\documentclass[twocolumn,usenatbib]{aastex62}

\usepackage{latexsym}
\usepackage{color}
\usepackage{amsmath}
\usepackage{amssymb}
\usepackage{graphicx}
\usepackage{aas_macros}
\usepackage{comment}

\graphicspath{{./}{figures/}}

\submitjournal{ApJL}

\shorttitle{Early galaxy formation with PBHs}
\shortauthors{Liu, B. \& Bromm, V.}

\begin{document}

\title{Accelerating early {massive} galaxy formation with primordial black holes}

\correspondingauthor{Boyuan Liu}
\email{boyuan@utexas.edu}

\author[0000-0002-4966-7450]{Boyuan Liu \href{https://orcid.org/0000-0002-4966-7450}{\includegraphics[width=2.5mm]{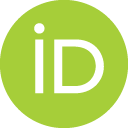}}}
\affiliation{Department of Astronomy, 
University of Texas at Austin, 2515 Speedway, Stop C1400, Austin, TX 78712, USA}
\affiliation{Institute of Astronomy, 
University of Cambridge, Madingley Road, Cambridge, CB3 0HA, UK}

\author[0000-0003-0212-2979]{Volker Bromm \href{https://orcid.org/0000-0003-0212-2979}{\includegraphics[width=2.5mm]{orcid.png}}}
\affiliation{Department of Astronomy, 
University of Texas at Austin, 2515 Speedway, Stop C1400, Austin, TX 78712, USA}

\begin{abstract}
Recent observations with JWST have identified several bright galaxy candidates at $z\gtrsim 10$, some of which appear unusually massive (up to $\sim 10^{11}\ \rm M_{\odot}$). Such early formation of massive galaxies is difficult to reconcile with standard $\Lambda\rm CDM$ predictions, demanding very high star formation efficiency (SFE), possibly even in excess of the cosmic baryon mass budget in collapsed structures. With an idealized analysis based on linear perturbation theory and the Press-Schechter formalism, we show that the observed massive galaxy candidates can be explained with lower SFE than required in $\Lambda\rm CDM$, if structure formation is {accelerated/seeded by massive ($\gtrsim 10^{9}\ \rm M_{\odot}$) primordial black holes (PBHs) that make a up a small fraction ($\sim 10^{-6}-10^{-3}$) of dark matter, considering existing empirical constraints on PBH parameters. We also 
discuss the potential observational signatures of PBH cosmologies in the JWST era. More work needs to be done to fully evaluate the viability of such PBH models to explain observations of the high-$z$ Universe.} 
\end{abstract}


\section{Introduction}
\label{s1}
Understanding the onset of star and galaxy formation at the end of the cosmic dark ages, a few 100 million years after the Big Bang, is one of the key goals of modern cosmology \citep[e.g.,][]{Barkana2001,Bromm2011}. With the successful launch of the James Webb Space Telescope (JWST), this formative period of cosmic history is now becoming accessible to direct observations, thus finally testing the theoretical framework of early structure formation. The latter extends the $\Lambda$CDM model, which is highly successful in accounting for galaxy formation and evolution following the epoch of reionization \citep{Springel2006, mo2010galaxy}, to the first billion years after the Big Bang.   

The initial JWST imaging via the Cosmic Evolution Early Release Science (CEERS) survey has discovered a population of surprisingly massive galaxy
candidates at $z\gtrsim 10$, with inferred stellar masses of $\gtrsim 10^9$\,M$_{\odot}$ \citep{Atek2022,Finkelstein2022,Harikane2022,Naidu2022,Yan2022}. The current record among these photometric detections reaches out to $z\simeq 16.7$  \citep{Donnan2022}. Such massive sources so early in cosmic history would be difficult to reconcile with the expectation from standard $\Lambda$CDM \citep{Boylan-Kolchin2022,Inayoshi2022,Lovell2022}, and this includes similarly over-massive (up to $\sim 10^{11}\ \rm M_{\odot}$) galaxy candidates detected at $z\simeq 10$ \citep{Labbe2022}. An important caveat here is that the early release JWST candidate galaxies are based on photometry only, rendering their redshift and spectral energy distribution (SED) fits uncertain \citep{Steinhardt2022}, until spectroscopic follow-up will become available. Part of the seeming discrepancy may also be alleviated by the absence of dust in sources at the highest redshifts, thus boosting their rest-frame UV luminosities \citep[e.g.,][]{Jaacks2018,Ferrara2022}.  

It is a long-standing question in cosmology when the first galaxies emerged, and how massive they were, going back to the idea that globular clusters formed at the Jeans scale under the conditions immediately following recombination \citep{Peebles1968}. The modern view of early structure formation, based on the dominant role of cold dark matter (CDM), posits initial building blocks for galaxy formation that are low-mass, of order $10^6$\,M$_{\odot}$, virializing at $z\simeq 20-30$ \citep{Couchman1986, Haiman1996}. 

Numerous models have
been proposed to suppress small-scale fluctuations, thus delaying the onset of galaxy formation to later times, in response to empirical hints for the lack or absence of low-mass objects that are otherwise predicted by CDM models \citep[][]{Bullock2017}. An extreme scenario here is the fuzzy dark matter (FDM) model, assuming ultra-light axion-like dark matter particles with corresponding de Broglie wavelengths of $\sim 1$\,kpc \citep{Hui2017}. Quantum pressure would thus prevent the collapse of structure on the scale of dwarf galaxies \citep[e.g.,][]{Sullivan2018}. 

The opposite effect, deriving models to {\it accelerate} early structure formation, beyond the $\Lambda$CDM baseline prediction, as may be required to explain the massive JWST galaxy candidates, is much more challenging. One recent study invokes the presence of an early dark energy (EDE) component, resulting in such an accelerated formation of high-redshifts structures \citep{Klypin2021}. {We here explore a different possibility of boosting the emergence of massive galaxies in early cosmic history with primordial black holes (PBHs), considering the isocurvature perturbations from PBHs that increase the power of density fluctuations in addition to the standard $\Lambda$CDM adiabatic mode \citep[i.e., the `Poisson' effect;][]{Carr2018}. In addition, we assess the scenario in which the most massive galaxies reported in \citet{Labbe2022} form in halos seeded by massive PBHs that are very rare and evolve in isolation at high $z$ \citep[i.e., the `seed' effect;][]{Carr2018}.}

\section{PBH structure formation}
\label{s2}

\begin{figure}[tbp]
    \centering
    \includegraphics[width=1\columnwidth]{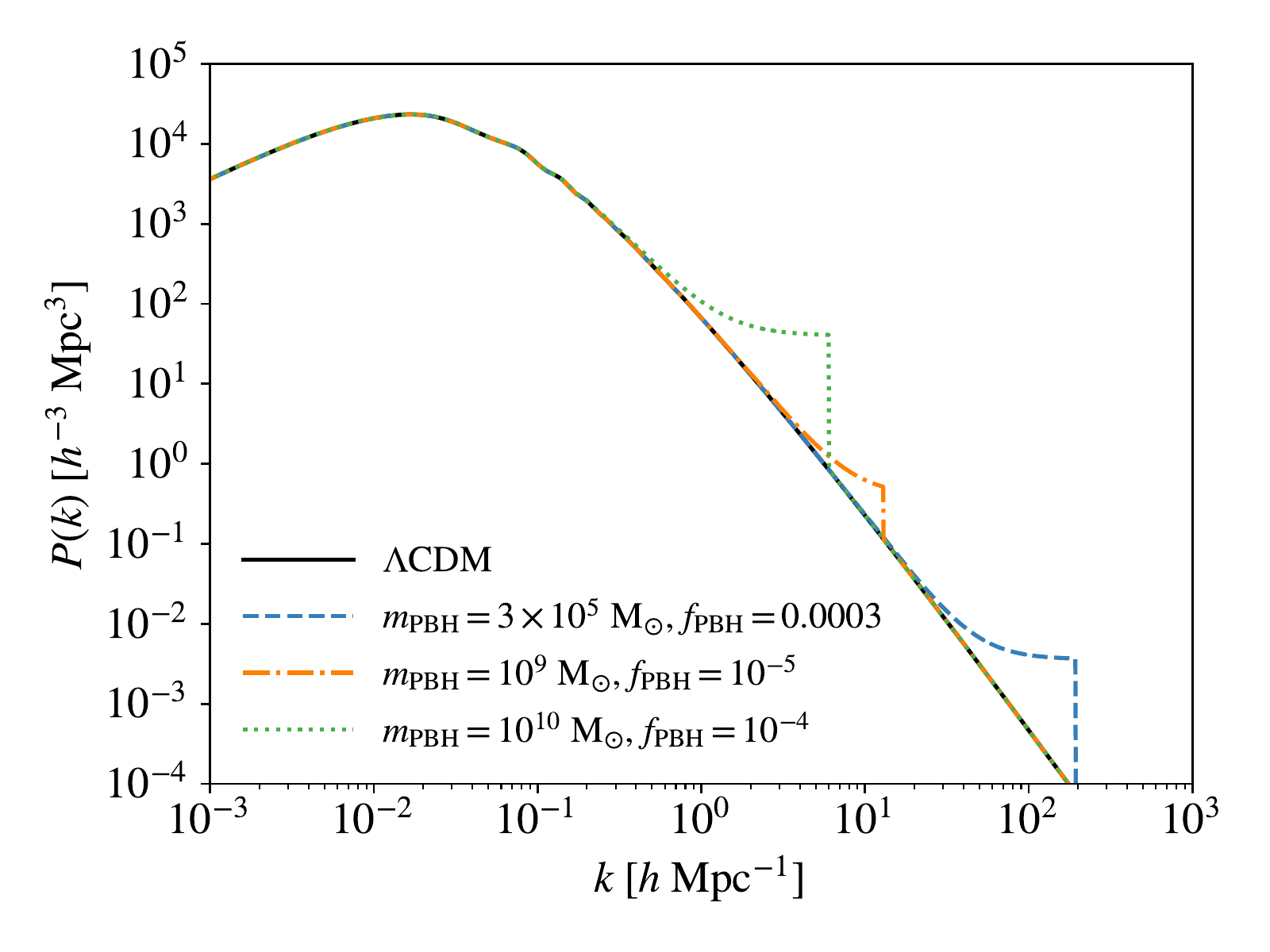}
    \vspace{-20pt}
    \caption{The power spectra of dark matter density field at $z=0$ in linear perturbation theory for 3 PBH models with {$(m_{\rm PBH}/{\rm M_{\odot}}, f_{\rm PBH})=(3\times 10^{-5},0.0003)$ (dashed), $(10^{9},10^{-5})$ (dashed-dotted) and $(10^{10},10^{-4})$ (dotted)}, compared with the standard $\Lambda$CDM power spectrum (solid) measured by \citet{PlanckCollaboration2020} from the \textsc{python} package \href{https://bdiemer.bitbucket.io/colossus/index.html}{\textsc{colossus}} \citep{colossus}.}
    \label{powspec}
\end{figure}

For simplicity, we adopt a monochromatic\footnote{The mass spectrum of PBHs can take a variety of forms from different formation mechanisms as discussed in, e.g., \citet{Carr2018,Tada2019,Carr2019,Carr2020,Carr2021qcd}. For observational constraints on PBHs, considering an extended mass spectrum of PBHs is a two-edged sword \citep{carr_primordial_2019}: On the one hand, this tends to make the constraints more stringent in terms of the maximum fraction of dark matter in PBHs from a given mass band \citep{Carr2017,Carr2021}. On the other hand, the total PBH density may suffice to explain all dark matter, even if the density in any particular mass band is small and within the observational bounds, as shown in \citet{Garcia-Bellido2019}. } mass function
for PBHs, such that a PBH model is specified by the BH mass $m_{\rm PBH}$ and fraction $f_{\rm PBH}$ of dark matter in the form of PBHs. {We start with the `Poisson' effect \citep{Carr2018} in which} PBHs produce isocurvature perturbations in the density field on top of the standard adiabatic mode due to the random distribution of PBHs (at small scales). These isocurvature perturbations only grow in the matter-dominated era. 
{As an extension of the formalism in \citet{Afshordi2003,Kashlinsky2016,Cappelluti2022}, where PBHs make up all dark matter ($f_{\rm PBH}\sim 1$), now treating $f_{\rm PBH}$ as a free parameter, the linear power spectrum (extrapolated to $a=1$) of dark matter density fluctuations can be written as \citep{Inman2019,Liu2022}}: 
\begin{align}
    P(k)&= P_{\rm ad}(k) + P_{\rm iso}(k)\ ,\notag\\
    P_{\rm iso}(k)&\simeq [f_{\rm PBH} D_{0}]^{2}/\bar{n}_{\rm PBH}\ ,\label{epowspec}
\end{align}
where $P_{\rm ad}(k)$ is the standard adiabatic mode in $\rm \Lambda CDM$ cosmology\footnote{We use the $\rm \Lambda CDM$ power spectrum (for the adiabatic mode) measured by \citet{PlanckCollaboration2020} with $\Omega_{\rm m}=0.3153$, $\Omega_{\rm b}=0.0493$, $h=0.6736$, $\sigma_{8}=0.8111$ and $n_{s}=0.9649$ from the \textsc{python} package \href{https://bdiemer.bitbucket.io/colossus/index.html}{\textsc{colossus}} \citep{colossus}.}, $\bar{n}_{\rm PBH}=f_{\rm PBH}\frac{3H_{0}^{2}}{8\pi G}(\Omega_{\rm m}-\Omega_{\rm b})/m_{\rm PBH}$ is the cosmic (co-moving) number density of PBHs, and $D_{0}$ is the growth factor of isocurvature perturbations evaluated at $a=1$, given by \citep{Inman2019}:
\begin{align}
    D(a)&\simeq \left(1+\frac{3\gamma}{2a_{-}}s\right)^{a_{-}}-1\ ,\quad s=\frac{a}{a_{\rm eq}}\ ,\notag\\
    \gamma&=\frac{\Omega_{\rm m}-\Omega_{\rm b}}{\Omega_{\rm m}}\ ,\quad a_{-}=\frac{1}{4}\left(\sqrt{1+24\gamma}-1\right)\ \label{dgrow}\ ,
\end{align}
where $a_{\rm eq}=1/(1+z_{\rm eq})$ is the scale factor at matter-radiation equality with $z_{\rm eq}\simeq 3400$. { So far, we have ignored the higher-order (non-linear) `seed' effect \citep{Carr1984,Carr2018}}, as well as mode mixing\footnote{At early stages when overdensities are very small, the isocurvature mode and adiabatic mode are uncorrelated. However, at later stages (e.g., $z\sim 10-20$, when the first galaxies form), the two modes can be mixed as PBHs follow the large-scale adiabatic mode to fall into larger structures and meanwhile induce/disrupt DM structures around themselves on small scales \citep{Liu2022}.}, lacking a self-consistent linear perturbation theory that can take them into account. {Such effects tend to enhance the perturbations induced by PBHs at large scales but suppress structure formation at small scales with non-linear dynamics around PBHs \citep{Liu2022}. Particularly, we expect linear perturbation theory to break down at a certain (mass) scale $M_{\rm bk}$, overproducing the abundance of structures below $M_{\rm bk}$. We hypothesize $M_{\rm bk}$ to be between $m_{\rm PBH}$ and the mass $M_{\rm B}(m_{\rm PBH},z)\sim [(z+1)a_{\rm eq}]^{-1}m_{\rm PBH}$ of overdense ($\delta\gtrsim 1$) regions bound to individual PBHs, when they evolve in isolation \citep{Carr2018}. In light of this, as a heuristic approach, we further impose a conservative cut-off to the isocurvature term in Eq.~\ref{epowspec} to suppress the power at scales smaller than $M_{\rm bk}\sim m_{\rm PBH}$:
\begin{align}
    P_{\rm iso}(k)=0\ ,\quad k>(2\pi^{2}\bar{n}_{\rm PBH}/f_{\rm PBH})^{-1/3}.\label{cut}
\end{align}
As examples, Fig.~\ref{powspec} shows the power spectra for three models that turn out to be representative in our analysis below (see Sec.~\ref{s3} and Table~\ref{model}). 
}

Once $P(k)$ is known, we use the Press-Schechter (PS) formalism \citep{Press_1974,mo2010galaxy} with a Gaussian window function to calculate the halo mass function (HMF), $dn(M,z)/dM$, including corrections for ellipsoidal dynamics \citep{Sheth1999}. Following \citet{Boylan-Kolchin2022}, given the star formation efficiency (SFE), $\epsilon\equiv M_{\star}/(f_{\rm b}M_{\rm halo})$, we then derive the (co-moving) cumulative stellar mass density contained within galaxies above a certain stellar mass $M_{\star}$ as
\begin{align}
    \rho_{\star}(>M_{\star},z)&=\epsilon f_{\rm b}\rho(>M_{\rm halo},z)\notag\\
    &=\epsilon f_{\rm b}\int_{M_{\rm halo}}^{\infty}M\frac{dn(M,z)}{dM}dM\ , \label{esmd}
\end{align}
where $f_{\rm b}=\Omega_{\rm b}/\Omega_{\rm m}$ is the cosmic average baryon fraction. 
Exploring whether PBH scenarios can explain the massive galaxy candidates reported by \citet{Labbe2022}, in the following sections we compare the $\rho_{\star}(>M_{\star},z)$ { and number density of massive galaxies} predicted by our PBH models with observations, and discuss the general signatures of PBHs observable by JWST.

\section{PBH signatures in the JWST era}
\label{s3}

\subsection{PBH models required to explain current JWST results}

\begin{figure}[tbp]
    \centering
    \includegraphics[width=1\columnwidth]{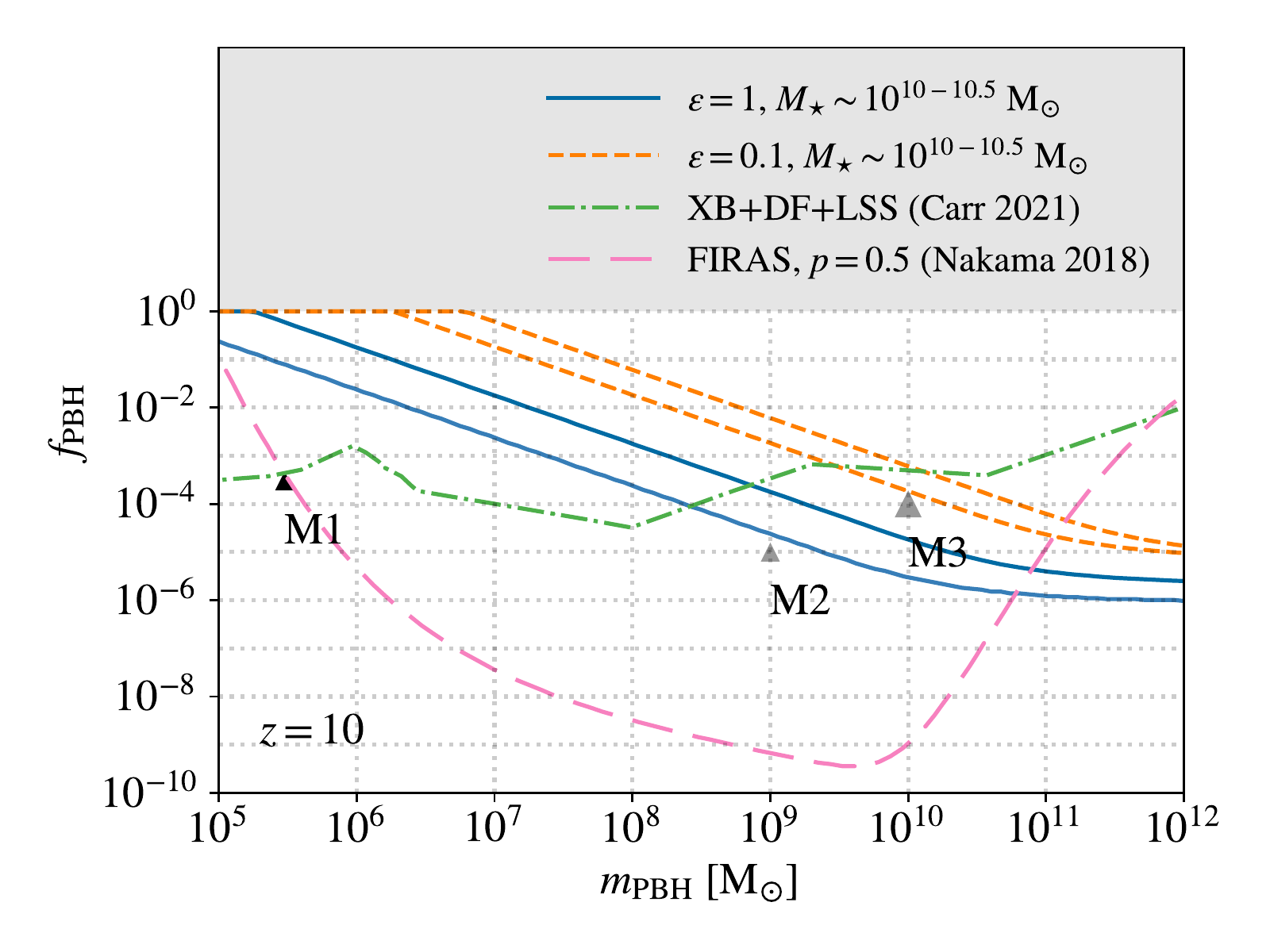}
    \vspace{-20pt}
    \caption{The PBH parameters required to explain the cumulative stellar mass density at $z\sim 10$ measured by JWST \citep{Labbe2022} { with the `Poisson' effect}. The PBH models consistent with JWST observations are shown with the solid and dashed curves for $\epsilon=1$ and 0.1, respectively. In each case, the upper (lower) line corresponds to the data point at the limiting mass $M_{\star}\sim 10^{10.5\ (10)}$. We also show the constraint from the FIRAS CMB $\mu$-distortion limit with a non-Gaussianity parameter $p=0.5$ \citep[long-dashed]{Nakama2018}, and the combined constraint from X-ray binaries \citep[XB,][]{Inoue2017}, dynamical friction \citep[DF,][]{Carr1999} and large-scale structures \citep[LSS,][]{Carr2018}, compiled by \citet[dashed-dotted]{Carr2021}. {Three sample models are labeled with triangles (see Table~\ref{model}), among which the big (small) triangle(s) can (not) explain the JWST results. The darker triangle (M1) satisfies both observational constraints while the fainter triangles (M2 and M3) only satisfy the XB+DF+LSS constraint.}}
    \label{pbh_pdg}
\end{figure}

Based on 14 galaxy candidates with masses $\sim 10^{9}-10^{11}\ \rm M_{\odot}$ at $7<z<11$, identified in the JWST CEERS program, \citet[see their fig.~4]{Labbe2022} derive the cumulative stellar mass density at $z=8$ and 10 for $M_{\star}\gtrsim 10^{10}\ \rm M_{\odot}$. In particular, they find $\rho_{\star}(\gtrsim 10^{10}\ \rm M_{\odot})\simeq 1.3_{-0.6}^{+1.1}\times 10^{6}\ \rm M_{\odot}\ Mpc^{-3}$ and $\rho_{\star}(\gtrsim 10^{10.5}\ \rm M_{\odot})\simeq 9_{-6}^{+11}\times 10^{5}\ \rm M_{\odot}\ Mpc^{-3}$ at $z\sim 10$, higher than the maximum achievable in $\Lambda\rm CDM$ (with $\epsilon=1$) by up to a factor of $\sim 50$. Using the formalism described in the previous section (Eqs.~\ref{epowspec}-\ref{esmd}), we find the PBH parameters that can reproduce these results for SFE values $\epsilon=1$ and 0.1, comparing with existing observational constraints in the $f_{\rm PBH}$-$m_{\rm PBH}$ space, 
as shown in Fig.~\ref{pbh_pdg}. 

It turns out that to explain the observational data at $M_{\star}\sim 10^{10.5\ (10)}\ \rm M_{\odot}$, we must have $m_{\rm PBH}f_{\rm PBH}\gtrsim 1.8\times 10^{5}\ (2.4\times 10^{4})\ \rm M_{\odot}$ for $\epsilon<1$ and $m_{\rm PBH}f_{\rm PBH}\gtrsim 6.1\times 10^{6}\ (1.8\times 10^{6})\ \rm M_{\odot}$ for $\epsilon<0.1$, {indicating that we need relatively massive PBHs with $m_{\rm PBH}\gtrsim 10^{5}\ \rm M_{\odot}$ since $f_{\rm PBH}\le 1$. To form such massive PBHs in the standard spherical collapse scenario \citep{Escriva2022}, very large perturbations are required out of inflation, e.g., from non-Gaussian tails produced by interacting quantum fields \citep[e.g.,][]{Frampton2016,Panagopoulos2019,Atal2019}, or oscillatory features in the inflationary power spectrum \citep{Carr2019}.} Such PBH models are strongly constrained by observations of $\mu$-distortion in the cosmic microwave background (CMB), if primordial density fluctuations are Gaussian \citep[see their figs.~10 and 16]{Carr2021}. 

Nevertheless, the constraints can be weaker when the Gaussian assumption is relaxed \citep[e.g.,][]{Nakama2018}. For instance, with the long-dashed curve in Fig.~\ref{pbh_pdg} we show a particular case of the phenomenological non-Gaussian model in \citet{Nakama2016} with $p=0.5$, where $p$ is the non-Gaussianity parameter ($p=2$ means Gaussian). In this case, PBHs with $m_{\rm PBH}\gtrsim 10^{11}\ \rm M_{\odot}$ can still have large enough $f_{\rm PBH}$ ($\gtrsim 10^{-5}$) to explain both the JWST observations and the CMB $\mu$-distortion limit \citep{Nakama2018}, as measured by the COBE Far Infrared Absolute Spectrophotometer (FIRAS). { However, such models are disfavored by the non-detection of BHs above $10^{11}\ \rm M_{\odot}$ \citep[except for Phoenix~A,][]{Brockamp2016}. Therefore, an even higher degree of non-Gaussianity is required to explain the JWST results with PBH models of $m_{\rm PBH}\lesssim 10^{11}\ \rm M_{\odot}$ and high enough $f_{\rm PBH}$. 
Alternatively}, the CMB $\mu$-distortion constraint can be evaded, if PBHs grow significantly between the $\mu$-distortion epoch ($7\times 10^{6}\ \mathrm{s}<t<3\times 10^{9}\ \rm s$) and matter-radiation equality \citep{Carr2021}{, or if PBHs form in non-standard scenarios such as inhomogeneous baryogenesis with the modified Affleck-Dine mechanism \citep{Kawasaki2019,Kasai2022}. }

Beyond the {$\mu$-distortion} constraint, PBHs with $m_{\rm PBH}\gtrsim 10^{5}\ \rm M_{\odot}$ are also constrained by X-ray binaries \citep[XB,][]{Inoue2017}, infall of PBHs into the Galactic center by dynamical friction \citep[DF,][]{Carr1999}, and large-scale structure statistics \citep[LSS,][]{Carr2018}, which together require {$f_{\rm PBH}\lesssim 10^{-4}-10^{-3}$ for $m_{\rm PBH}\sim 10^{5}-10^{11}\ \rm M_{\odot}$} (see the dashed-dotted curve in Fig.~\ref{pbh_pdg}). Such constraints are generally weaker, allowing a PBH abundance for $m_{\rm PBH}\gtrsim 10^{9}\ \rm M_{\odot}$ sufficient to produce the high stellar mass density in massive galaxies at $z\sim 10$, inferred by JWST \citep{Labbe2022}. {However, in this regime ($f_{\rm PBH}\lesssim 10^{-3}$), the `seed' effect likely dominates at $z\gtrsim 10$ according to the criterion that the fraction of mass bound to PBHs in the universe is $\ll 1$, i.e., $f_{\rm PBH}\ll (1+z)a_{\rm eq}$ \citep{Carr2018}. Besides, isocurvature perturbations purely from the `Poisson' effect (without the cut-off in Eq.~\ref{cut}) are strongly constrained by high-$z$ observations \citep[e.g.,][]{Sekiguchi2014,Murgia2019,Tashiro2021}, such that PBH models with $m_{\rm PBH}f_{\rm PBH}\gtrsim 170\ \rm M_{\odot}$ (and $f_{\rm PBH}>0.05$) are ruled out by high-$z$ Lyman-$\alpha$ forest data \citep{Murgia2019}. This motivates us to further explore the `seed' effect. }

\begin{figure}
    \centering
    \includegraphics[width=\columnwidth]{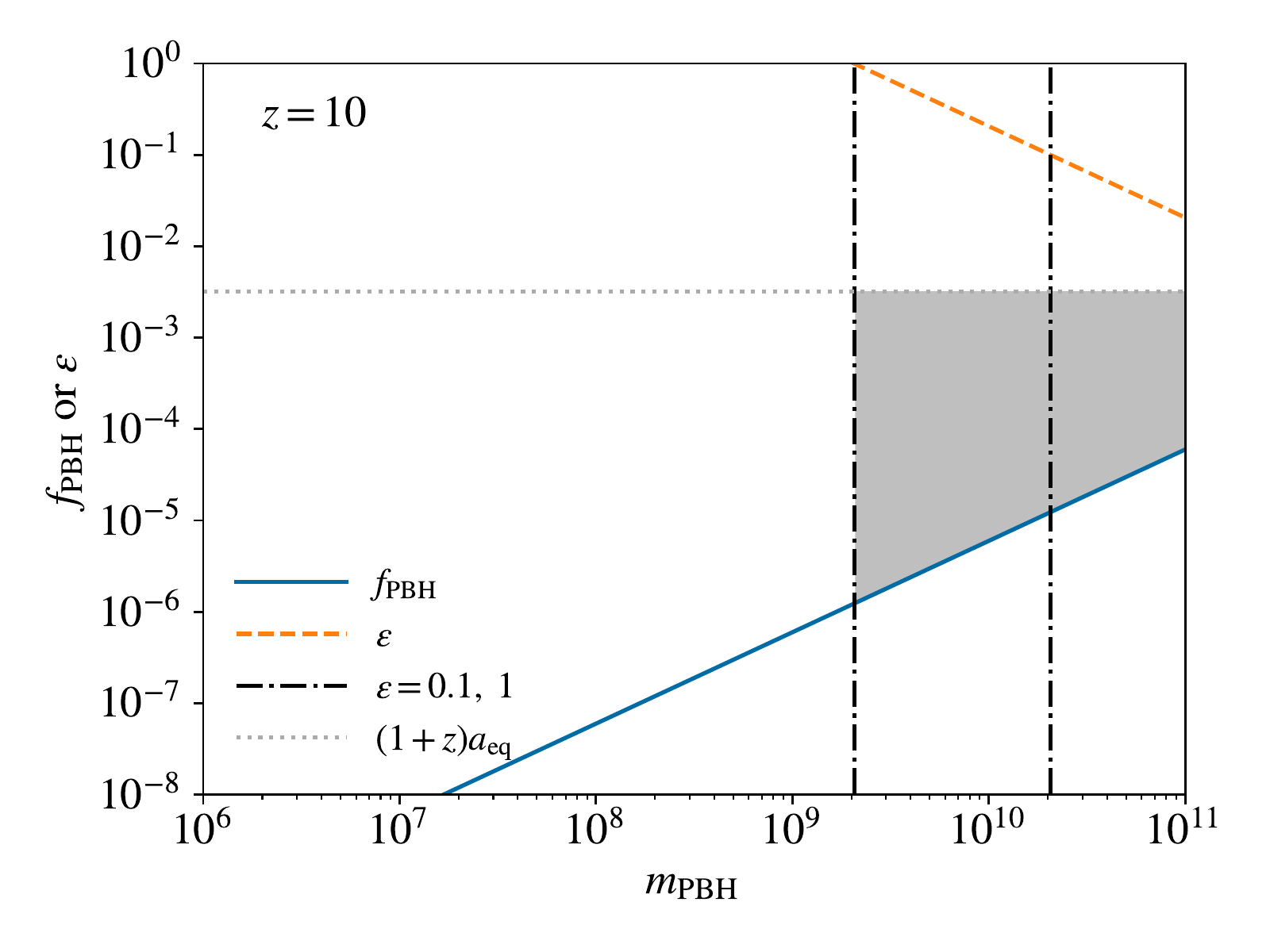}
    \vspace{-20pt}
    \caption{The PBH parameters (shaded region) required to explain the (co-moving) number density $n_{\rm g}\sim 2\times 10^{-5}\ \rm Mpc^{-3}$ of observed massive ($M_{\star}\sim 10^{11}\ \rm M_{\odot}$) galaxies at $z\sim 10$ \citep{Labbe2022,Boylan-Kolchin2022} with the `seed' effect. The shaded region is defined by simple arguments of number counts, baryon mass budget and dominance of the `seed' effect: 
    $\bar{n}_{\rm PBH}>n_{\rm g}$ (solid), $\epsilon\equiv M_{\star}/[f_{\rm b}M_{\rm B}(m_{\rm PBH},z)]<1$ (and 0.1, see the dashed and dashed-dotted lines) and $f_{\rm PBH}<(1+z)a_{\rm eq}$ \citep[dotted, which is the general criterion for the assumption that PBHs evolve in isolation to hold,][]{Carr2018}. }
    \label{pbh}
\end{figure}

{ 
We perform an idealized calculation to explore what PBH models are needed to form the most massive galaxies observed by JWST purely with the `seed' effect. 
Assuming that the observed massive galaxies form in halos seeded by PBHs that grow in isolation as $M_{\rm halo}\sim M_{\rm B}(m_{\rm PBH},z)\sim [(z+1)a_{\rm eq}]^{-1}m_{\rm PBH}$ \citep{Carr2018}, we require that (i) the average (co-moving) number density $\bar{n}_{\rm PBH}$ of PBHs is larger than that of the observed massive galaxies $n_{\rm g}\sim 2\times 10^{-5}\ \rm Mpc^{-3}$ \citep{Labbe2022,Boylan-Kolchin2022}, and that (ii) the PBH-seeded halos have enough gas to form $M_{\star}\sim 10^{11}\ \rm M_{\odot}$ of stars, i.e., $\epsilon f_{\rm b}M_{\rm B}=M_{\star}$ can be satisfied given the limit of $\epsilon$. As shown in Fig.~\ref{pbh}, we find that the `seed' effect can explain the JWST results with generally less extreme PBH models of $m_{\rm PBH}f_{\rm PBH}\gtrsim 3\times 10^{5\ (3)}\ \rm M_{\odot}$ and $m_{\rm PBH}\gtrsim 2\times 10^{10\ (9)}\ \rm M_{\odot}$ for $\epsilon\lesssim 0.1\ (1)$. Moreover, considering the non-linear dynamics around PBH-seeded halos, it is found in cosmological simulations that structures smaller than $\sim M_{\rm B}\ (\gtrsim 10^{12}\ \rm M_{\odot}$ at $z\lesssim 10$ for $m_{\rm PBH}\gtrsim 10^{9}\ \rm M_{\odot}$) are significantly less abundant than predicted by the `Poisson' effect \citep[see their fig.~14]{Liu2022}. This can weaken/lift the Lyman-$\alpha$ forest constraint that is sensitive for $k\sim 0.7-10\ h\rm\ Mpc^{-1}$ \citep{Murgia2019}, corresponding to $M_{\rm halo}\sim 10^{9}-10^{12}\ \rm M_{\odot}$.
}

\subsection{Signatures of representative PBH models}

{To further demonstrate the effects of PBHs in high-$z$ galaxy/structure formation potentially observable by JWST, we focus on 3 models representative for typical regions in the PBH parameter space (Fig.~\ref{pbh_pdg}), as listed in Table~\ref{model}. Here M1 satisfies all observational constraints considered above, while M2 and M3 only satisfy the XB+DF+LSS combined constraint. On the other hand, for the `Poisson' effect, M3 is consistent with the recent JWST results in \citet{Labbe2022} with $\epsilon\sim 0.1-1$, while M1 and M2 cannot reproduce the observations even with $\epsilon=1$. For the `seed' effect, M2 can also marginally explain the JWST observations. } 

\begin{table}[htbp]
    \centering
    \vspace{-10pt}
    \caption{Representative PBH models {(see Fig.~\ref{powspec} for the corresponding linear power spectra of density fluctuations)}.}
    \begin{tabular}{cccc}
    \hline
        Model & $m_{\rm PBH}\ [\rm M_{\odot}]$ & $f_{\rm PBH}$ & $m_{\rm PBH}f_{\rm PBH}\ [\rm M_{\odot}]$ \\
    \hline
        M1 & $3\times 10^{5}$ & $0.0003$ & 90\\
        M2 & $10^{9}$ & $10^{-5}$ & 10000\\
        M3 & $10^{10}$ & $10^{-4}$ & $10^{6}$\\
    \hline
    \end{tabular}
    \vspace{-10pt}
    \label{model}
\end{table}

\begin{figure}[htbp]
    \centering
    \includegraphics[width=0.95\columnwidth]{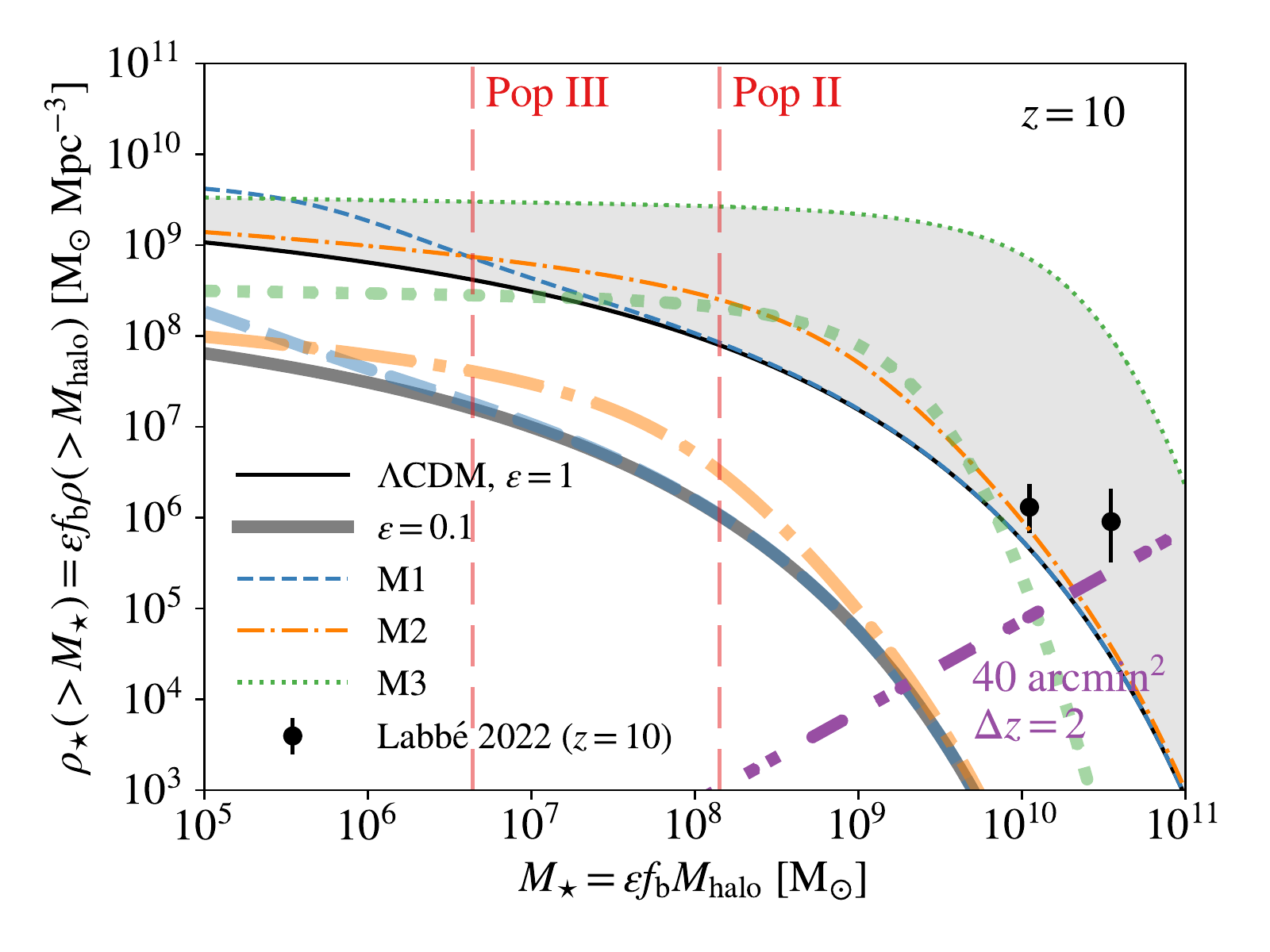}
    \vspace{-20pt}\\
    \includegraphics[width=0.95\columnwidth]{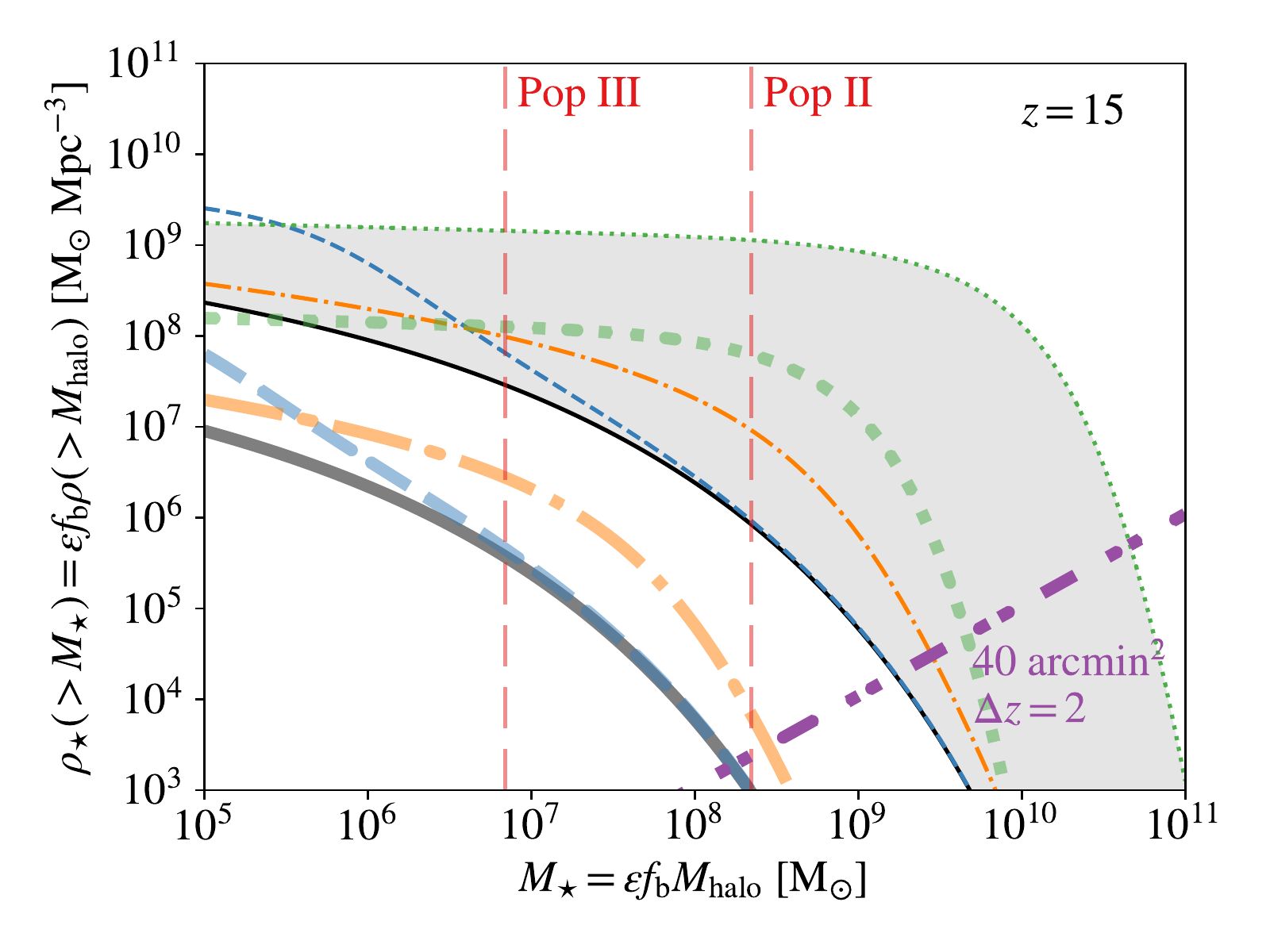}
    \vspace{-20pt}\\
    \includegraphics[width=0.95\columnwidth]{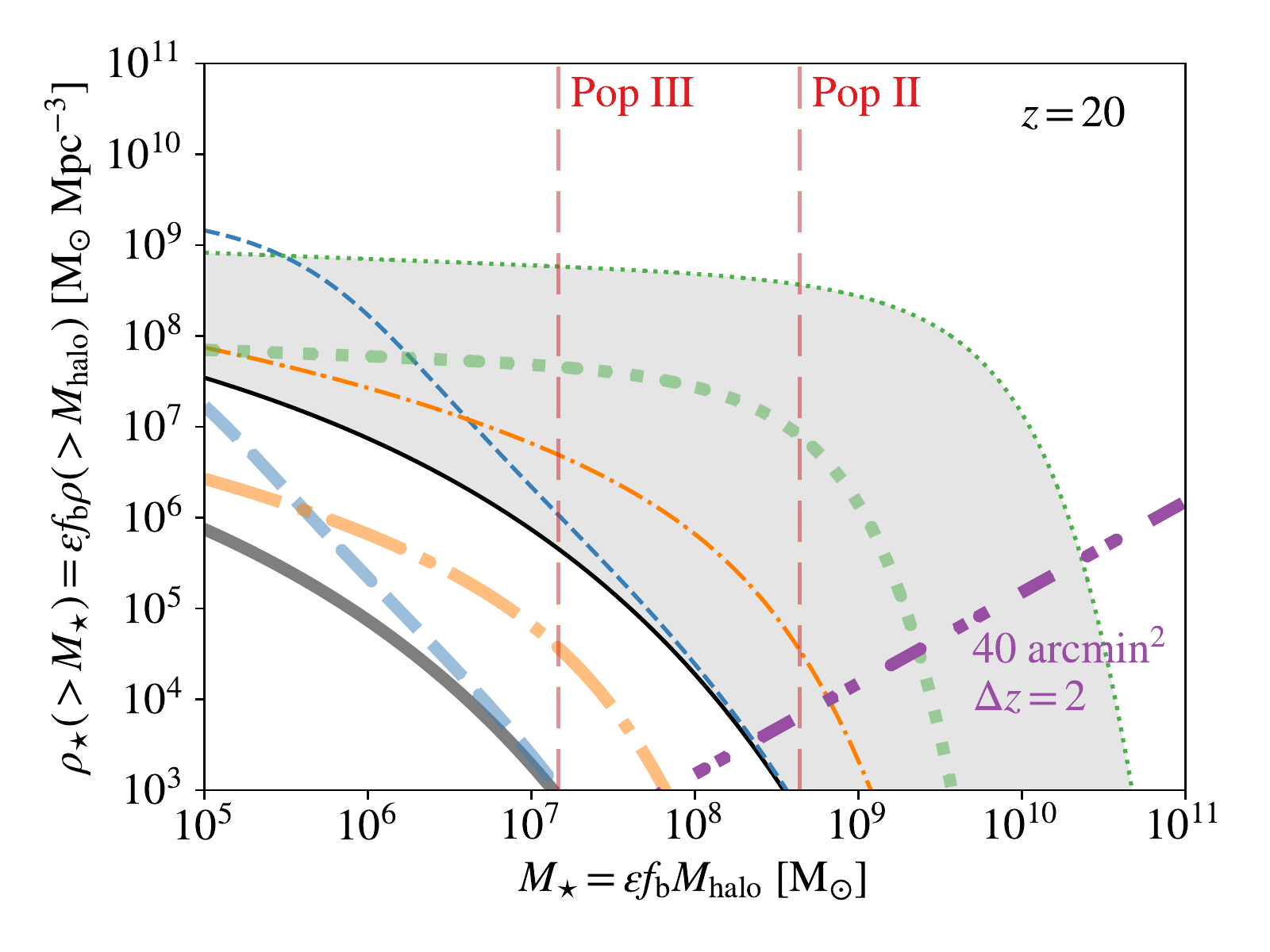}
    \vspace{-10pt}
    \caption{The cumulative (co-moving) stellar mass density in galaxies more massive than $M_{\star}$ at $z=10$ (top), 15 (middle) and 20 (bottom). {The results for standard $\Lambda$CDM, PBH models M1, M2 and M3 (see Table~\ref{model} and Fig.~\ref{pbh_pdg}) are shown with the solid, dashed, dashed-dotted and dotted curves for $\epsilon=1$ (thin) and 0.1 (thick).} The shaded region can only be populated by galaxies in PBH cosmologies. 
    The results inferred from the recent JWST observations at $z\sim 10$ \citep{Labbe2022} are denoted by the data points with error bars. The long-dashed vertical lines mark the lower mass limits of Pop~III and II galaxies to be detected by JWST CEERS NIRCam LW imaging (see the main text). We also plot the number count limit $M_{\star}/V_{\rm com}$ with the dashed-dotted-dotted line, given the co-moving volume $V_{\rm com}$ for a survey like CEERS of $40\ \rm arcmin^2$ and $\Delta z=2$.}
    
    \label{smd}
\end{figure}

In Fig.~\ref{smd}, we present the results for $\rho_{\star}(> M_{\star})$ in standard $\Lambda\rm CDM$ (solid), M1 (dashed), M2 (dashed-dotted) and M3 (dotted) for $\epsilon=1$ (thin) and 0.1 (thick) at $z=10$, 15 and 20. 
We also plot the lower mass limits of galaxies detectable by JWST CEERS NIRCam LW imaging (for an exposure time of $\sim 2000\ \rm s$) in Fig.~\ref{smd}, considering a magnitude limit $m_{\rm F444W}\sim 28$ for the filter F444W with a signal-to-noise ratio $\rm (SNR)\gtrsim 5$, derived from the JWST Exposure Time Calculator\footnote{\url{https://jwst.etc.stsci.edu/}}. We calculate the stellar mass limit as the total mass of stars formed in a star formation event whose peak luminosity matches the magnitude limit according to the stellar population synthesis code \href{https://www.astro.uu.se/~ez/yggdrasil/yggdrasil.html}{\textsc{yggdrasil}} \citep{Zackrisson2011}. For Population~III (Pop~III) stars we adopt their (instantaneous-burst) Pop III.1 model with an extremely top-heavy Salpeter initial mass function (IMF) in the range of $50-500\rm\ M_{\odot}$, based on \citet{Schaerer2002}, while for Population~II (Pop~II) stars we use their $Z=0.0004$ model from Starburst99 \citep{Leitherer1999,Vazquez2005} with a universal \citet{Kroupa2001} IMF in the interval $0.1-100\ \rm M_{\odot}$. For both Pop~III and II, we consider a nebula covering fraction of $f_{\rm cov}=0.5$ and no Lyman-$\alpha$ transmission. Note that these limits will be reduced by about one order of magnitude for future deeper surveys with longer ($\sim 100\ \rm h$) exposures \citep{Zackrisson2012}.

As shown in Fig.~\ref{smd}, $\rho_{\star}(> M_{\star})$ can be significantly increased by PBHs and the effects are {the strongest for halos of $M_{\rm halo}\sim m_{\rm PBH}-10m_{\rm PBH}$, and stronger at higher $z$}. In the extreme model M3, the stellar mass budget at $z\lesssim 20$ will be dominated by galaxies more massive than $ 10^{8}\ \rm M_{\odot}$ given $\epsilon\gtrsim 0.1$. {Fully taking into account the `seed' effect may further increase the mass budget of massive halos ($M_{\rm halo}\sim m_{\rm PBH}-M_{\rm B}$) with respect to that of the smaller ones \citep{Liu2022}.} Even in M1, $\rho_{\star}(> M_{\star})$ is higher by a factor of $\gtrsim 10$ than in $\Lambda\rm CDM$ for $M_{\star}\sim 10^{5}\ \rm M_{\odot}$ at $z\gtrsim 15$. Considering the area ($\sim 40\ \rm arcmin^{2}$) of CEERS, corresponding to the number count limit in Fig.~\ref{smd} (i.e., the dashed-dotted-dotted line), we conclude that the fact that JWST has detected galaxies above $10^{8}\ \rm M_{\odot}$ at $z\gtrsim 15$ \citep[e.g.,][]{Atek2022,Donnan2022,Harikane2022,Naidu2022,Yan2022} already requires very high SFE, $\epsilon\gtrsim 0.1$, for Pop~II star formation in $\Lambda\rm CDM$, consistent with the UV luminosity function analysis by \citet{Inayoshi2022}. On the other hand, for structure formation accelerated by PBHs, such galaxies can also form in more massive halos with lower SFE (down to $0.01$ for M3). Finally, our calculation indicates that at $z\gtrsim 20$ detection of any Pop~II galaxy by a survey like CEERS is impossible in $\Lambda\rm CDM$, and even Pop~III galaxies require $\epsilon\gtrsim 0.1$ to be detected with stellar masses above $10^{7}\ \rm M_{\odot}$, {which is inconsistent with the theoretical and observational upper limits, $M_{\rm PopIII}\lesssim 10^{6}\ \rm M_{\odot}$, on the total mass of (active) Pop~III stars a halo can host \citep{Yajima2017,Bhatawdekar2021}.} However, with massive PBHs like those in M3, it is possible to detect Pop~II galaxies as massive as $\gtrsim 4\times 10^{8}\ \rm M_{\odot}$ with $\epsilon\gtrsim 0.01$.

\section{Summary and discussion}
\label{s4}

The recent detection of surprisingly massive ($\sim 10^{10}-10^{11}\ \rm M_{\odot}$) galaxy candidates at $z\sim 10$ by JWST \citep{Labbe2022}, if confirmed, brings new challenges to $\Lambda\rm CDM$ (and a broad range of dynamical dark energy models) which cannot provide enough baryonic matter in collapsed structures for such early massive galaxy formation \citep{Boylan-Kolchin2022,Lovell2022,Menci2022}. Note that the source properties derived from photometry are sensitive to the underlying SED fitting templates, such that the stellar masses can be lower by up to 1.6~dex than reported in \citet{Labbe2022}, if other templates with improved physical justification are adopted, removing the tension with $\Lambda\rm CDM$ \citep[see their fig.~3]{Steinhardt2022}. Currently no template can well match the observed photometry self-consistently, and follow-up spectroscopic observations are needed to robustly pin down {the nature and properties of these galaxy candidates \citep{Furlanetto2022}}.

Assuming that the results in \citet{Labbe2022} are true {and considering only the `Poisson' effect of PBHs, we use a simple analytical model based on linear perturbation theory and the PS formalism to show that such early formation of massive galaxies is possible if PBHs make up part of dark matter with $m_{\rm PBH}f_{\rm PBH}\gtrsim 6\times 10^{6}\ (2\times 10^{5})\ \rm M_{\odot}$ for SFE $\epsilon<0.1\ (1)$. The `seed' effect, on the other hand, requires $m_{\rm PBH}f_{\rm PBH}\gtrsim 3\times 10^{5\ (3)}\ \rm M_{\odot}$ and $m_{\rm PBH}\gtrsim 2\times 10^{10\ (9)}\ \rm M_{\odot}$ for $\epsilon\lesssim 0.1\ (1)$. Such massive PBHs are mostly ruled out by observations of the CMB $\mu$-distortion, although} this strong constraint relies on the assumptions that primordial density fluctuations are Gaussian and that PBHs (formed in the standard scenario of primordial density fluctuations) hardly grow during the radiation-dominated era. { Besides, strong isocurvature perturbations purely from the `Poisson' effect of PBHs with $m_{\rm PBH}f_{\rm PBH}\gtrsim 170\ \rm M_{\odot}$ at scales of $k\sim 0.7-10\ h\rm\ Mpc^{-1}$ are ruled out by high-$z$ Lyman-$\alpha$ forest data \citep{Murgia2019}. Nevertheless, considering the non-linear dynamics around massive ($m_{\rm PBH}\gtrsim 10^{9}\ \rm M_{\odot}$) PBHs, the abundance of halos at such scales can be much lower than predicted by the `Poisson' effect \citep{Liu2022}, lifting the Lyman-$\alpha$ forest constraint. The $\mu$-distortion constraint can also be evaded by relaxing the underling assumptions or considering none-standard PBH formation mechanisms \citep[e.g.,][]{Kawasaki2019,Kasai2022}. In this way,} the other constraints from X-ray binaries \citep{Inayoshi2022}, dynamical friction \citep{Carr1999} and large-scale structures \citep{Carr2018} allow a substantial region in the PBH parameter space with $m_{\rm PBH}\sim 10^{9}-10^{11}\ \rm M_{\odot}$ and {$f_{\rm PBH}\sim 10^{-6}-10^{-3}$} that is consistent with the recent JWST observations (and no detection of BHs above $10^{11}\ \rm M_{\odot}$). 

{ However, considering that PBHs can grow by up to two orders of magnitude through the acquisition of dark matter halos by $z\sim 10$ with optimistic spherical\footnote{The growth of PBH masses can be much weaker with advection dominated disk accretion, e.g., up to $\sim 2\%$ for $m_{\rm PBH}\lesssim 10^{7}\ \rm M_{\odot}$ at $z\gtrsim20$ \citep{Hasinger2020}.} accretion \citep[e.g.,][]{Mack2007,DeLuca2020}, and the non-detection of BHs above $\sim 10^{11}\ \rm M_{\odot}$, the allowed region will further shrink. 
On the other hand, if the stellar masses measured by \citet{Labbe2022} are overestimated (or some of the galaxy candidates are actually at lower $z$) so that the stellar mass density at $z\sim 10$ is lower in reality, less extreme PBH models would be able to explain them for the same value of $\epsilon$, and the same PBH model could allow lower values of $\epsilon$. For instance, if the inferred stellar masses were indeed reduced by 1.6 dex \citep{Steinhardt2022} in follow-up observations, we would only need $m_{\rm PBH}f_{\rm PBH}\gtrsim 2\times 10^{5}\ \rm M_{\odot}$ to form the observed galaxies with $\epsilon<0.025$ from the `Poisson' effect, while the `seed' effect requires $m_{\rm PBH}f_{\rm PBH}\gtrsim 200\ (2)\ \rm M_{\odot}$ and $m_{\rm PBH}\gtrsim 5\times 10^{8\ (7)}\ \rm M_{\odot}$ for $\epsilon\lesssim 0.1\ (1)$. }

We also find that the effects of PBHs are stronger at higher $z$, 
implying that stronger signatures of PBHs may be found in future wider and deeper surveys by JWST. Actually, if the object CEERS-1749 reported in \citet{Naidu2022} is a galaxy at $z\sim 17$, its large mass ($\sim 5\times 10^{9}\ \rm M_{\odot}$) is also in $>3\sigma$ tension with $\Lambda\rm CDM$ \citep[see their fig.~6]{Lovell2022}. Even if follow-up studies do not find such excess of very massive galaxies at higher $z$, this may not necessarily rule out our PBH models, since the SFE can evolve rapidly with redshift at Cosmic Dawn ($z\sim 6-30$) due to metal enrichment, the build-up of radiation backgrounds and the transition in dominant stellar population \citep[see e.g.,][]{Fialkov2019,Mirocha2019,Schauer2019,Liu2020}. 
{Besides, if the average SFE is not significantly lower with PBHs, the accelerated structure formation leads to accelerated star formation that can also facilitate cosmic reionization. Interestingly, this may explain the recent observations of Lyman-$\alpha$ emitting galaxies that support a double-reionization scenario with the first full ionization event happening at $z\sim 10$ \citep{Salvador-Sole2017,Salvador-Sole2022}. Very massive ($\sim 100-10^{3}\ \rm M_{\odot}$) Pop~III stars are required to produce this double-reionization feature in $\Lambda\rm CDM$ \citep{Salvador-Sole2017}, while less extreme stellar populations may be sufficient with accelerated star formation by PBHs given proper values of SFE.}

{Note that the PBH masses and enhancement of density fluctuations by PBHs required to explain the JWST observations in our case ($m_{\rm PBH}\gtrsim 10^{9}\ \rm M_{\odot}$ and $m_{\rm PBH}f_{\rm PBH}\gtrsim 3\times 10^{3}\ \rm M_{\odot}$) are significantly higher (and working at larger scales) than those considered by previous studies for stellar-mass PBHs \citep[$m_{\rm PBH}\lesssim 100\ \rm M_{\odot}$ and $m_{\rm PBH}f_{\rm PBH}\lesssim 30\ \rm M_{\odot}$, see, e.g.,][]{Kashlinsky2016,Gong2017,Inman2019,Cappelluti2022,Liu2022}. In these studies, the standard picture of first star formation in molecular-cooling minihalos remains unchanged but with earlier onset and increased abundances of star-forming halos. However, in our extreme PBH models, star formation may first occur in atomic-cooling halos ($M_{\rm halo}\gtrsim 10^{8}\ \rm M_{\odot}$), seeded by PBHs or formed in a top-down fashion by fragmentation of massive ($\gtrsim 10^{11}\ \rm M_{\odot}$) structures around PBHs. The accretion feedback from BHs can also significantly affect nearby formation of stars and direct collapse black holes \citep[e.g.,][]{Pandey2018,Aykutalp2020,Liu2022}. Considering the gravitational, hydrodynamic and radiative effects, early star formation in the presence of massive PBHs can leave strong imprints in the CMB, 21-cm signal, reionization, cosmic infrared and X-ray backgrounds \citep[see, e.g.,][]{Sekiguchi2014,Kashlinsky2016,Gong2017,Murgia2019,Hasinger2020,Tashiro2021,Cappelluti2022,Minoda2022}. These important aspects involving non-linear dynamics and baryonic physics are beyond the scope of our exploratory work. Follow-up studies with more detailed modeling of the interactions between PBHs, baryons and (non-PBH) dark matter (in cosmological simulations) are required to fully understand the roles played by massive PBHs in structure/galaxy/star formation and evaluate the viability of such PBH models.} 

In general, the recent discovery of potentially very early massive galaxy formation by JWST \citep{Labbe2022} hints at faster structure formation in the high-$z$ universe, above the $\Lambda\rm CDM$ baseline \citep{Boylan-Kolchin2022,Lovell2022}, which can be achieved if {a small fraction ($\sim 10^{-6}-10^{-3}$)} of dark matter is composed of massive ($\gtrsim 10^{9}\ \rm M_{\odot}$) PBHs{, although more work needs to be done to check whether such fast structure formation is consistent with other observations of the high-$z$ Universe}. A similar trend is also seen in observations of (proto-) galaxy clusters that show an excess of strong-lensing sources \citep{Meneghetti2020,Meneghetti2022} and of star formation \citep{Remus2022} that are difficult to explain in $\Lambda \rm CDM$. Strikingly, this trend for accelerated structure formation at high-$z$ goes in the opposite direction of invoking the suppression of fluctuations to account for the well-known small-scale problems of $\Lambda\rm CDM$. Together with other hints for PBHs \citep[see e.g.,][]{Clesse2018}, these findings imply that the nature of dark matter may be more complex than our standard expectation, e.g., involving important sub-components such as PBHs. This challenge calls for the thorough theoretical exploration of the alternatives to $\Lambda \rm CDM$, in conjunction with advanced observational campaigns to probe the high-redshift universe. Here, we are entering an exciting period of discovery, with frontier facilities, such as JWST, Euclid, the Square Kilometre Array (SKA), as well as the Einstein Telescope (ET) and Laser Interferometer Space Antenna (LISA) gravitational wave observatories.




\section*{Acknowledgments}
We would like to thank Antonio Riotto and the anonymous referee for their helpful comments.

\newpage

\bibliographystyle{aasjournal}
\bibliography{ref}


\label{lastpage}
\end{document}